\documentclass{aa}  
\usepackage{color}
\usepackage{graphicx}
\usepackage{lscape}
\usepackage{natbib}
\usepackage{natbib,twoopt}
\usepackage[varg]{txfonts}
\usepackage{url}
\usepackage{xspace}
\usepackage{amssymb}
\usepackage{stmaryrd}
\usepackage{siunitx}
\usepackage{textcomp}
\usepackage{pifont}

\usepackage[colorlinks = true,
linkcolor= blue,
citecolor = blue,
filecolor = blue,
urlcolor = blue,
]{hyperref}

\bibpunct{(}{)}{;}{a}{}{,}

\def\etal{{et\,al.}\ }

\newcommand{\pn}{\object{IPHASX\,J055226.2$+$323724}\xspace}

\newcommand{\cmark}{\ding{51}}%
\newcommand{\xmark}{\ding{55}}%

\newcounter{Rco}

\newcommand{\logg}{\mbox{$\log g$}\xspace}
\newcommand{\loggw}[1]{\mbox{$\log g\hspace{-0.5mm} =\hspace{-0.5mm}  #1$}}

\newcommand{\Teff}{\mbox{$T_\mathrm{eff}$}\xspace}

\newcommand{\Lsol}{$L_\odot$}
\newcommand{\Msol}{$M_\odot$}
\newcommand{\Rsol}{$R_\odot$}

\begin{document}

\title{The unusual planetary nebula nucleus in the Galactic open cluster M37 and six further hot white dwarf candidates}

\author{Klaus Werner\inst{1} \and Nicole Reindl\inst{2,}\inst{3} \and Roberto Raddi\inst{4} \and
Massimo Griggio\inst{5,} \inst{6} \and 
Luigi R. Bedin \inst{7} \and 
Mar\'ia E. Camisassa\inst{4}
\and \\
Alberto Rebassa-Mansergas\inst{4,}\inst{7} \and 
Santiago Torres\inst{4,}\inst{7} \and 
Peter Goodhew\inst{8}}

\institute{Institut f\"ur Astronomie und Astrophysik, Kepler Center for
  Astro and Particle Physics, Eberhard Karls Universit\"at, Sand~1, 72076
  T\"ubingen, Germany\\ \email{werner@astro.uni-tuebingen.de} 
\and
  Landessternwarte Heidelberg, Zentrum f\"ur Astronomie, Ruprecht-Karls-Universit\"at, Königstuhl~12, 69117 Heidelberg, Germany  
\and
  Institut f\"ur Physik und Astronomie, Universit\"at Potsdam, Karl-Liebknecht-Stra\ss e 24/25, 14476, Potsdam, Germany
\and
Departament
de F\'isica, Universitat Polit\`ecnica de Catalunya, c/Esteve Terrades 5, E-08860 Castelldefels, Spain
\and
Dipartimento di Fisica, 
Università di Ferrara, 
Via Giuseppe Saragat 1, 
Ferrara I-44122, Italy
\and
INAF - Osservatorio Astronomico di Padova, 
Vicolo dell’Osservatorio 5, 
Padova I-35122, 
Italy
\and
Institute for Space Studies of Catalonia, 
c/Gran Capit\`a 2--4, 
Edif. Nexus 104, 
08034 Barcelona, 
Spain
\and
W4 3EQ London, UK
}

\date{Received 17 June 2023 / Accepted 15 September 2023}

\authorrunning{K. Werner \etal}
\titlerunning{Hot white dwarf candidates in M37}

\abstract{
Planetary nebulae in Galactic open star clusters are rare objects; only three are known to date. They are of particular interest because their distance can be determined with high accuracy, allowing one to characterize the physical properties of the planetary nebula and its ionizing central star with high confidence. Here we present the first quantitative spectroscopic analysis of a central star in an open cluster, namely the faint nucleus of \pn\ in M37. This cluster contains 14 confirmed white dwarf members, which were previously used to study the initial-to-final-mass relation of white dwarfs, and six additional white dwarf candidates. We performed an atmosphere modeling of spectra taken with the 10m Gran Telescopio Canarias. The central star is a hot hydrogen-deficient white dwarf with an effective temperature of 90\,000\,K and spectral type PG1159 (helium- and carbon-rich). We know it is about to transform into a helium-rich DO white dwarf because the relatively low atmospheric carbon abundance indicates ongoing gravitational settling of heavy elements. The star belongs to a group of hot white dwarfs that exhibit ultrahigh-excitation spectral lines possibly emerging from shock-heated material in a magnetosphere. We find a relatively high stellar mass of $M= 0.85^{+0.13}_{-0.14}$~\Msol. This young white dwarf is important for the semi-empirical initial-final mass relation because any uncertainty related to white-dwarf cooling theory is insignificant with respect to the pre-white-dwarf timescale. Its post-asymptotic-giant-branch age of $170\,000-480\,000$~yr suggests that the extended planetary nebula is extraordinarily old. We also performed a spectroscopic analysis of the six other white dwarf candidates of M37, confirming one as a cluster member.
}


\maketitle
%
%________________________________________________________________

\section{Introduction}
\label{sect:intro}

In a recent astro-photometric study of the Galactic open cluster M37 (NGC 2099), \cite{2022MNRAS.515.1841G} identified seven hot white dwarfs as cluster member candidates. One of these white dwarfs (labeled WD1 in their list) is a high-probability cluster member as well as the probable central star of a planetary nebula \citep[PN;][]{2020A&A...638A.103C}. The large, evolved, and bipolar PN \pn\ (or PN G177.5$+$03) was confirmed to be a member of M37 \citep{2022ApJ...935L..35F}. It is only the third known PN in a Galactic open cluster, the other two being the PNe in NGC\,6067 and Andrews-Lindsay~1
\citep{2022Galax..10...44F,2019MNRAS.484.3078F,2011MNRAS.413.1835P}.

The central star WD1 ({\it Gaia} DR3 3451205783698632704, RA $05^{\rm h}52^{\rm m}26^{\rm s}.19$, Dec. $+32^\circ37'24''.89$, J2000) in M37 is a faint (V = 19.16) blue star located at the geometrical center of the PN \citep{2022ApJ...935L..35F}. Low-resolution spectroscopy presented by \cite{2022MNRAS.515.1841G} shows it to be a hydrogen-deficient white dwarf with a spectral type between DO and PG1159 and an effective temperature above \Teff = 60\,000\,K.

The open cluster M37 has long been known to contain a large white dwarf population of at least 14 confirmed members \citep{kalirai2005,cummings2015}, including a very massive one (1.28\,M$_\odot$) whose progenitor evolution is uncertain \citep[][]{cummings2016}. Due to this richness of white dwarf members, M37 has been the focus of detailed studies on the initial-to-final-mass relation (IFMR) of white dwarfs that connects their masses to those of their progenitors \citep[][]{cummings2015, 2018ApJ...866...21C}.

\begin{figure*}
 \centering  \includegraphics[width=0.9\textwidth]{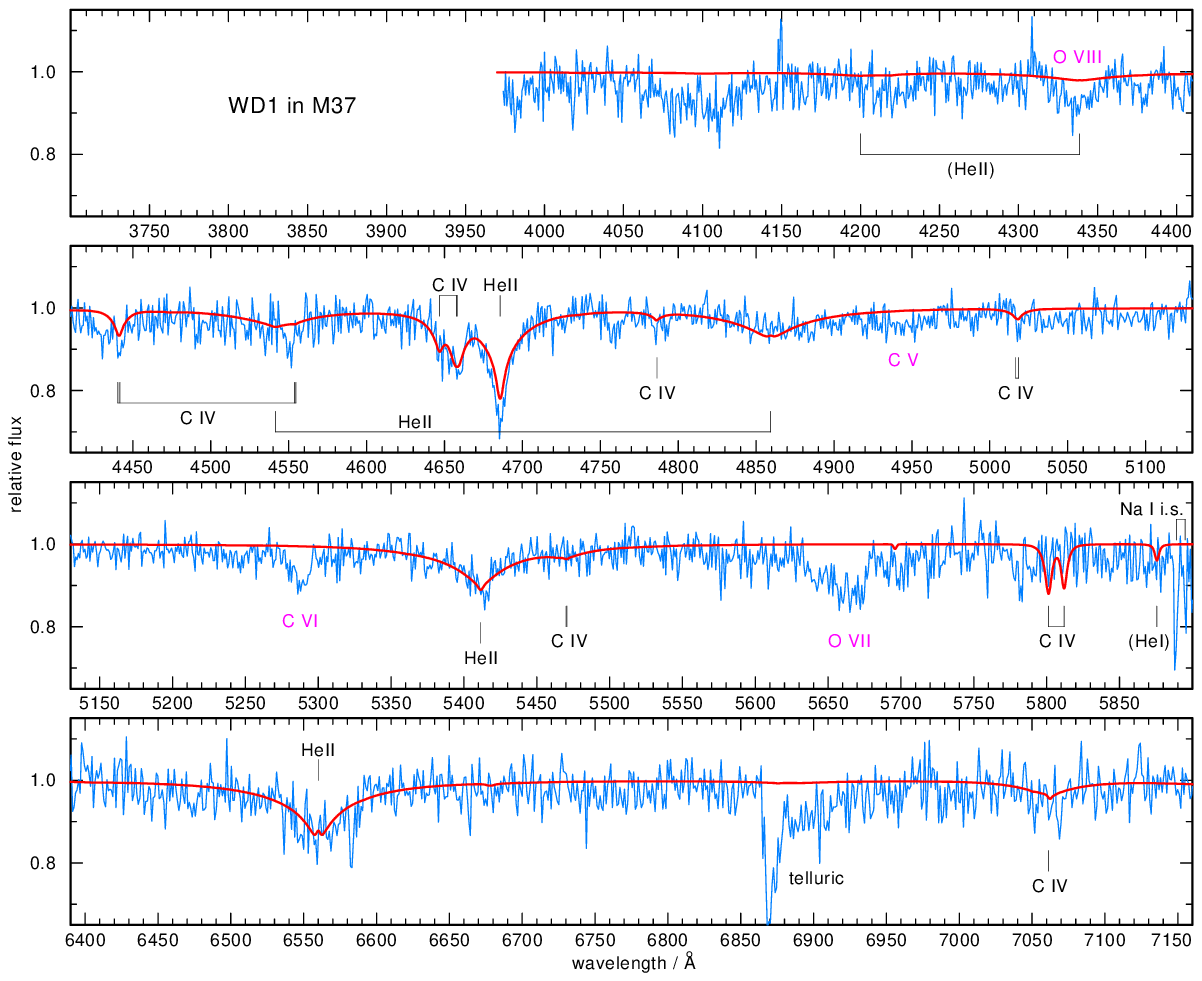}
 \caption{GTC spectrum of WD1, the hydrogen-deficient central star of the PN \pn, and our final model (red graph) with \Teff
   = 90\,000\,K, \logg = 8.3, He = 0.94, and C = 0.06 (mass
   fractions). Identified lines are labeled. Non-photospheric
   UHE lines of \ion{C}{v-vi} and \ion{O}{vii-viii} have violet labels.  }
\label{fig:wd1_fit}
\end{figure*}

\begin{table*}
    \centering
        \caption{Logs of the GTC observations.}
    \begin{tabular}{@{}lccccccl@{}}
\hline
\noalign{\smallskip}   
Name ID & Short Name & {\em Gaia} $G$  [mag] & Grism & Resolution [\AA] & Date & Seeing [arcsec] & Exposure [s] \\
\noalign{\smallskip}   
\hline
\noalign{\smallskip}   
WD1  &0552+3237  & 19.154 & R2000B   & 3.0 & 20220921   &   0.8   &   $4 \times 750$\\
     &           &        & R2500R   & 3.6 & 20220923   &   0.8   &   $4 \times 750$\\
WD2  &0552+3236  & 20.137 & R1000B   & 9.0 & 20221003   &   1.1   &   $2 \times 1400$\\
WD3  &0552+3231  & 19.769 & R1000B   & 9.0 & 20220929   &   1.1   &   $2 \times 1400$\\
WD4  &0551+3216  & 20.560 & R1000B   & 9.0 & 20221016   &   0.7   &   $4 \times 1000$\\
WD5  &0553+3229  & 20.660 & R1000B   & 9.0 & 20220929   &   1.2   &   $2 \times 1000$\\
     &           &        &          &     & 20221017   &   1.0   &   $2 \times 1000$\\
WD6  &0547+3246  & 19.917 & R1000B   & 9.0 & 20221003   &   0.7   &   $2 \times 1400$\\
WD7  &0548+3323  & 20.599 & R1000B   & 9.0 & 20220929   &   1.3   &   $2 \times 1000$\\
     &           &        &          &     & 20221016   &   0.8   &   $2 \times 1000$\\
\noalign{\smallskip}   
\hline
    \end{tabular}
    \label{tab:logs}
\end{table*}

Here we present new spectroscopic observations of WD1 and a non-local thermodynamic equilibrium (non-LTE) model atmosphere analysis as well as new deep imaging of the PN. In addition, we performed spectroscopic observations and analyses of the six other hot DA white dwarf member candidates of M37 identified by \cite{2022MNRAS.515.1841G} in order to further investigate their possible membership. In Sect.\,\ref{sect:observations} we introduce our observations. We continue in Sect.\,\ref{sect:analysis} with the spectral analysis of the white dwarfs, focusing on the PN nucleus WD1. In Sect.\,\ref{sect:sed} we perform fits to the spectral energy distribution (SED) of the white dwarfs and present the results. We finish in Sect.\,\ref{sect:discussion} with a summary of our study as well as a discussion on the contribution of the confirmed members to the IFMR of white dwarfs in open clusters.

\section{Observations}
\label{sect:observations}

\begin{figure*}
 \centering  \includegraphics[width=\columnwidth]{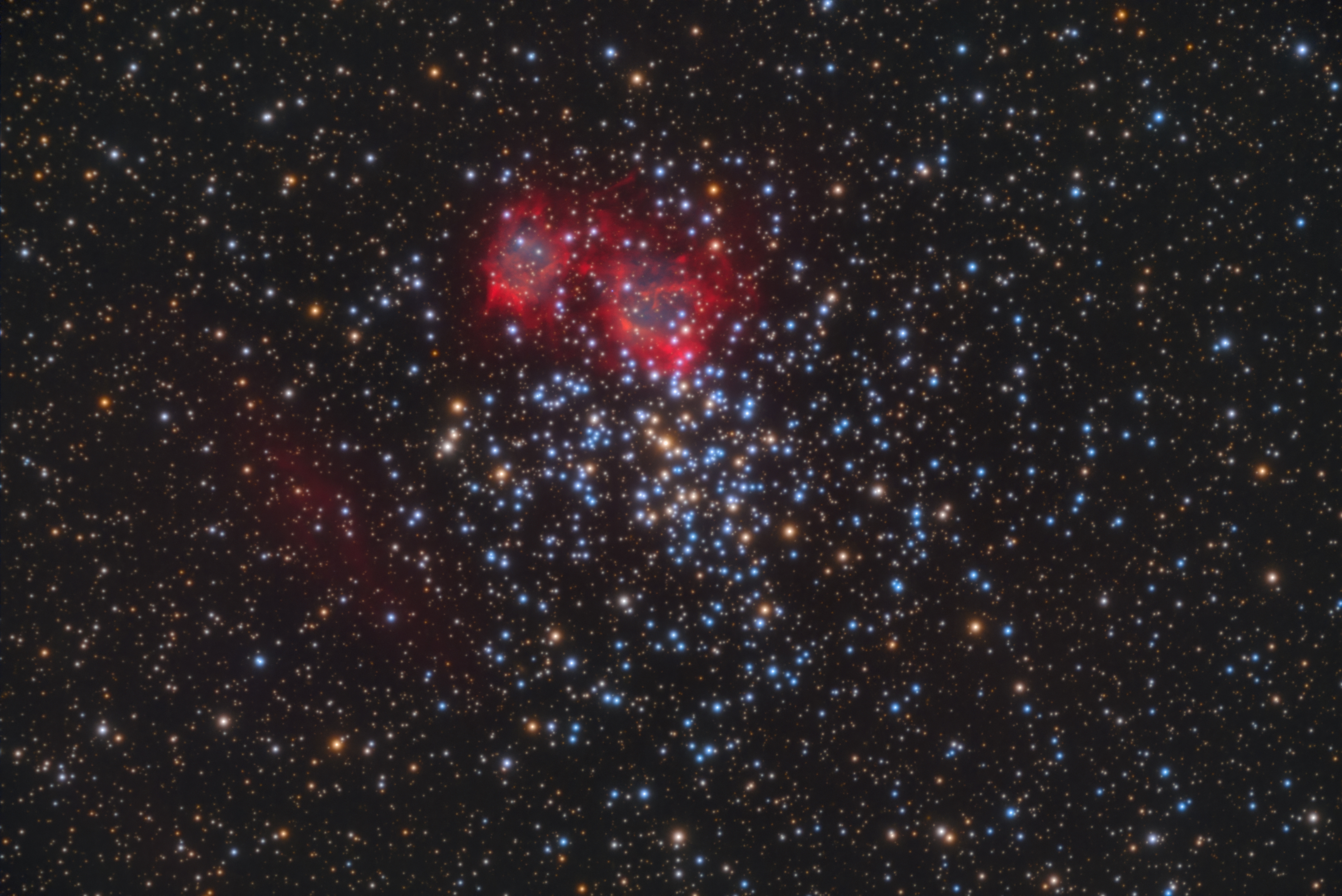}
 \centering  \includegraphics[width=\columnwidth]{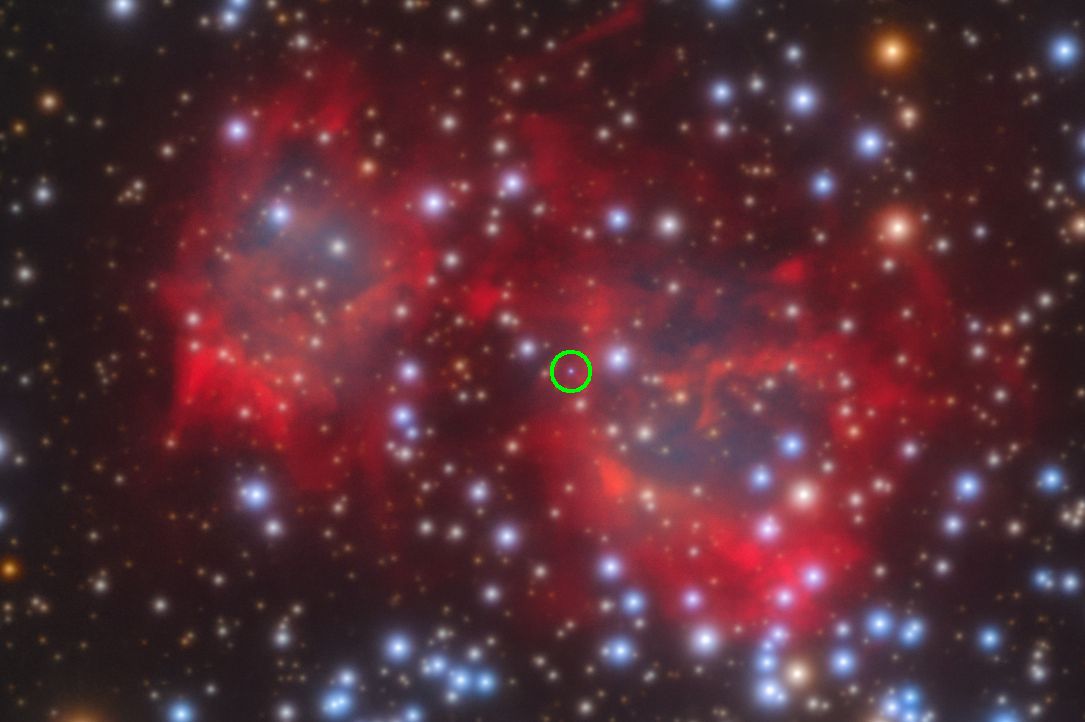}
 \caption{Image of the PN \pn in the open cluster M37. \emph{Left:} 
 Co-addition of all observations, with a total exposure time of 3.52 days as described in the text. H$\alpha$ is mapped to red and [\ion{O}{iii}] 5007\,\AA\ to green. Broadband frames of the stellar field are mapped to red, green, and blue. The height of the frame is $25'$. North is up and east to the left. Note the faint nebulosity at the southeastern rim of the cluster. \emph{Right:} Zoomed-in view of the left panel, with the PN central star circled in green.}
\label{fig:PN}
\end{figure*}

\subsection{Spectroscopy}

We acquired low-resolution long-slit optical spectra for all seven white dwarf candidates in M37 that were identified by Griggio and collaborators. These observations were performed in service mode at the 10m Gran Telescopio Canarias (GTC) in gray and dark time, in September and October  2022 (proposal ID: GTC72-22B).  We obtained $4\times 750$\,s exposures for WD1, both with the R2000B and R2500R gratings, employing a 1-arcsecond slit, and standard $2\times2$ binning and slow readout. The resolving power at the central wavelengths of both spectral regions is $R \sim 1500$. The other targets were observed with the R1000B grating, delivering a resolving power of $R \sim 800$, with exposure times ranging between 1000--1400\,s and obtaining between 2--4 exposures per star, depending on their apparent magnitudes. The average seeing was below 1.3 arcsec with no cloud coverage. All targets were observed at an airmass smaller than 1.3. The observing log is listed in Table\,\ref{tab:logs}.

We performed pre-reduction data manipulation, bias and flat-field correction, and optical extraction of the long-slit spectra \protect{\citep{horne1986}} by using the {\sc starlink} \protect{\citep{berry2022}} and {\sc pamela} software \protect{\citep{marsh1989}}. The wavelength and flux calibration used the nightly observations of HgAr and Ne lamps and the flux standards, respectively; finally, each sub-exposure was co-added, achieving a signal-to-noise ratio S/N\,$ \gtrsim 20$ at the H$\alpha$ wavelength. These tasks were performed using the {\sc molly} software \protect{\citep{marsh1989}. The rectified spectrum of WD1 is shown in Fig.\,\ref{fig:wd1_fit}.

\subsection{Imaging}

The PN \pn was discovered by \cite{2008PhDT.......499S} on images from the IPHAS survey (\citealt{2005MNRAS.362..753D}; see in particular the image versions presented by \citealt{2022ApJ...935L..35F} and \citealt{2022MNRAS.515.1841G}). The PN has a bipolar structure with a major axis of about 7.5 arcmin and a patchy internal structure. 

We present in Fig.\,\ref{fig:PN} our new image, which reveals more details of the PN structure. The data for the image were recorded on August 30 and September 10, 2022, using two remote robotic observatories located at the e-EyE telescope hosting site\footnote{\url{https://www.e-eye.es/en/hosting/}} near Fregenal de la Sierra in southwestern Spain. The one observatory was equipped with twin APM TMB LZOS 152 refractors\footnote{\url{https://www.apm-telescopes.net/en}} and QSI 6120wsg CCD cameras\footnote{\url{https://qsimaging.com/}}. Broadband filters used were Astrodon\footnote{\url{https://farpointastro.com/collections/astrodon}} red, green, and blue. Narrowband filters were Astrodon \ion{O}{iii} $\lambda 5007$\,\AA\ with 30\,\AA\ bandpass and Astrodon H$\alpha$ with 50\,\AA\ bandpass. The other observatory was equipped with a Celestron EdgeHD 14-inch reflector\footnote{\url{https://www.celestron.com/}} and a ZWO ASI6200MM Pro CCD camera\footnote{\url{https://astronomy-imaging-camera.com/}}. Employed filters were Chroma\footnote{\url{https://www.chroma.com/}} \ion{O}{iii} $\lambda 5007$\,\AA\ and H$\alpha$ both with 30\,\AA\ bandpass. The total \ion{O}{iii} and H$\alpha$ data integrations were 48:10 and 27:45 hours, respectively. The total integration time for broadband blue and green was 2:45 hours each and 2:55 hours for red. Image preprocessing and stacking was performed using {\sc PixInsight}\footnote{\url{https://pixinsight.com/}} and image processing using {\sc PixInsight} and {\sc Photoshop}\footnote{\url{https://www.adobe.com/de/products/photoshopfamily.html}}.

\section{Spectral analysis}
\label{sect:analysis}

\subsection{WD1, the central star of IPHASX\,J055226.2$+$323724}

\begin{figure*}
\begin{center}
\includegraphics[width=0.8\linewidth]{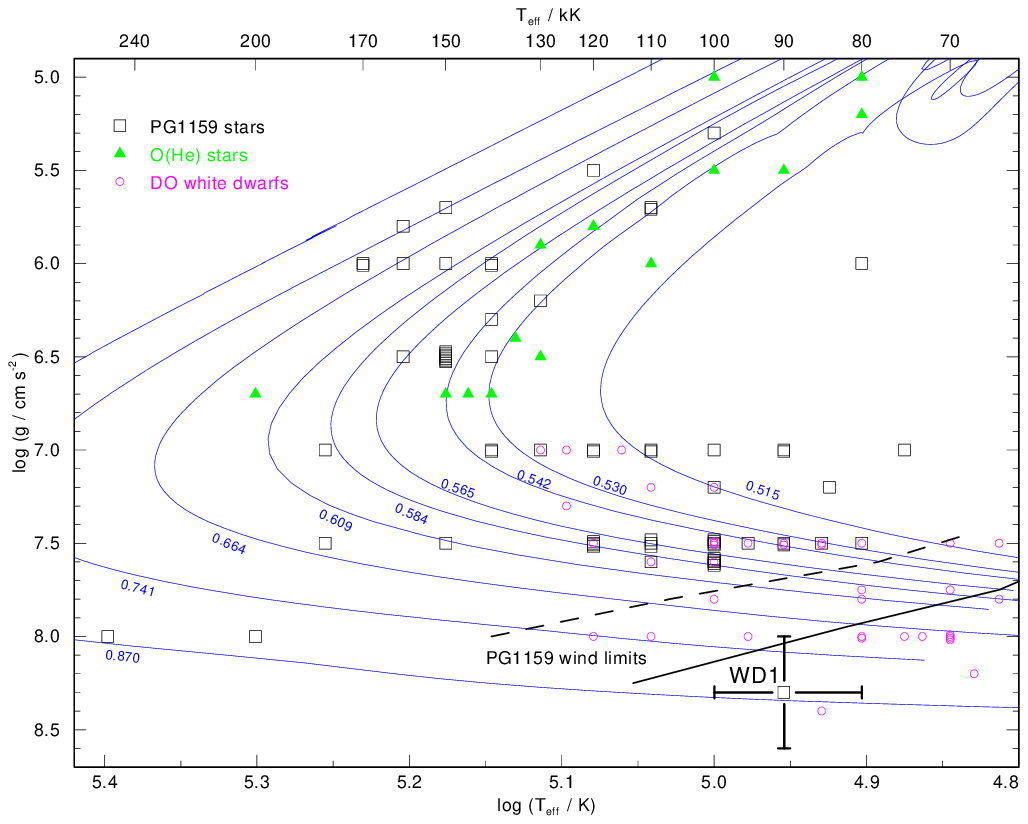}
\caption{Position and error bars of the PG1159 star WD1 in the Kiel
  diagram together with all known objects of this spectral class. Also shown are
 the positions of other hot helium-dominated objects, namely DO white
  dwarfs and O(He) stars. Evolutionary tracks by
  \citet{2009ApJ...704.1605A} are labeled with the stellar mass in
  solar units. The black line indicates the PG1159 wind limit
  according to \citet{2000A&A...359.1042U}, meaning that the mass-loss
  rate of the radiation-driven wind at this position of the evolutionary tracks
  becomes so weak that gravitational settling of heavy elements
  is able to remove the heavy elements from the white dwarf atmosphere. Thus, no PG1159
  stars should be found at significantly cooler temperatures. The dashed line is the
  wind limit assuming a mass-loss rate that is ten times lower.
  The fact that many PG1159 stars and DO white dwarfs are located at (half-)integer 
  \logg values is an artifact due to the spacing of model atmosphere grids employed for spectral analyses.
}
\label{fig:gteff}
\end{center}
\end{figure*}

The spectrum of WD1 is dominated by the hallmark of PG1159 stars, a
broad absorption trough formed by \ion{He}{ii} 4686\,\AA\ and a few
\ion{C}{iv} lines around 4650\,\AA\ (Fig.\,\ref{fig:wd1_fit}). PG1159
stars are hydrogen-deficient objects that have suffered a late
helium-shell flash \citep[see, e.g.,][]{2006PASP..118..183W}.  WD1
belongs to the spectral subtype ``A'' \citep{1992LNP...401..273W},
indicating that the lines in the trough are in pure absorption and
that the effective temperature is around 90\,000\,K. The strong
decrement of the \ion{He}{ii} Pickering line series points at a
surface gravity larger than \logg = 7. In addition, a few other very
weak \ion{C}{iv} lines are detected. We also identify four
non-photospheric, ultrahigh-excitation (UHE) lines from \ion{C}{v},
\ion{C}{vi}, \ion{O}{vii}, and \ion{O}{viii}. These lines require
extremely high temperatures in their formation region \citep[on the
  order of $10^6$~K,][]{1995A&A...293L..75W}. About 10\%
of the hottest white dwarfs show this UHE phenomenon, and it was suggested that the UHE lines form in a circumstellar
magnetosphere \citep{2019MNRAS.482L..93R,2021A&A...647A.184R}.

For the spectral analysis of WD1 we computed a small grid of non-LTE
model atmospheres of the type introduced by
\citet{2014A&A...564A..53W}. It was calculated using the T\"ubingen
Model-Atmosphere Package (TMAP) for non-LTE plane-parallel models in
radiative and hydrostatic equilibrium \citep{2003ASPC..288...31W}. The
constituents of the models are helium and carbon. The grid covers the
range \Teff = 70\,000--110\,000\,K in steps of 10\,000\,K, and \logg =
7.5--8.5 in steps down to 0.2\,dex. At the outset the mass fractions of helium and
carbon were fixed to He = 0.91 and C = 0.09. This is a low carbon abundance for a PG1159 star and it was chosen because of the relative weakness of the \ion{C}{iv} lines compared to many other stars of the PG1159 class. In the course of the analysis, the carbon
abundance in some models was varied to find the finally adopted value and to
estimate errors. It should be noted that we do not attempt to model the non-photospheric UHE lines.

The best-fit model was determined by the following procedure. The
\ion{He}{ii} 4860\,\AA\ line is virtually absent in the observation
and it was found that within the considered temperature range  only
models with sufficiently high surface gravity (around \logg = 8.5) can
match the observation. A lower limit to the temperature is provided by
the absence of \ion{He}{i}~5876\,\AA\ in the observation, meaning that
\Teff\ is at least 80\,000\,K. An upper limit of 100\,000\,K was
estimated, because at higher values \ion{He}{ii} 4686\,\AA\ becomes
very weak and the \ion{He}{ii} Pickering lines at 5410 and 6560\,\AA\ develop
emission cores that are not observed. By visual inspection of the
helium lines we finally adopted \Teff $= 90\,000\pm10\,000$\,K and \logg
$= 8.3\pm0.3$. The carbon abundance is C $= 0.06\pm0.03$ by mass. The
final model spectrum is depicted in Fig.\,\ref{fig:wd1_fit}. The atmospheric parameters of WD1 together with other characteristics are summarized in Table~\ref{tab:resultsall}.

\begin{table}[t]
\begin{center}
\caption{Atmospheric properties and other parameters of the white dwarf central star WD1 in M37.
\tablefootmark{a} }
\label{tab:resultsall}
\begin{tabular}{rr}
\hline 
\hline 
\noalign{\smallskip}
Parameter                 & Value                 \\
%\noalign{\smallskip}
\hline
\noalign{\smallskip}
Spectral type             & PG1159                \\
\Teff/\,K                 & $90\,000 \pm 10\,000$ \\
$\log$($g$\,/\,cm\,s$^{-2}$) & $8.3 \pm 0.3$      \\
He                        & $0.94 \pm 0.03$       \\   
C                         & $0.06 \pm 0.03$       \\ 
%$V$ magnitude             & 19.16                 \\
%$E(B-V)$                  & 0.26                  \\
$L$\,/\,\Lsol             & $13^{+20}_{-11}$   \\
%$L$\,/\,\Lsol             & $8^{+11}_{-5}$   \\
\noalign{\smallskip}
$R$\,/\,\Rsol             & $0.015^{+0.011}_{-0.005}$\\
%$R$\,/\,\Rsol             & $0.0116^{+0.0052}_{-0.0028}$\\
\noalign{\smallskip}
$M$\,/\,\Msol\ (VLTP)     & $0.85^{+0.13}_{-0.14}$\\
%\noalign{\smallskip}
%$E(B-V)$\,/ mag           & 0.26                  \\
\noalign{\smallskip}
$d$ / pc (SED fit)  & $800^{+380}_{-250}$   \\
\noalign{\smallskip}
$d$ / pc ({\it Gaia} parallax)  & $1272^{+905}_{-418}$  \\
\noalign{\smallskip}
$d$ / pc (M37)            & $1490\pm90$\\         
\noalign{\smallskip}
\hline
\end{tabular} 
\tablefoot{  \tablefoottext{a}{Element abundances given in mass
    fractions. Stellar mass derived from VLTP tracks
    (Fig.\,\ref{fig:gteff}). The {\it Gaia} distance of WD1 is taken from
    \cite{2021AJ....161..147B} and the distance of M37 from \cite{2022MNRAS.511.4702G} using {\it Gaia} Early Data Release 3 data. }  } 
\end{center}
\end{table}

It can be seen
that the computed \ion{He}{ii} 4686\,\AA\ line is not deep enough. This could be related
to the \ion{He}{ii} line problem observed in other UHE stars
\citep{2021A&A...647A.184R}.  This means that the \ion{He}{ii} lines are
too deep to be fitted by any model. On the other hand, this problem
also affects other lines besides \ion{He}{ii} 4686\,\AA,\ but in the
case of WD1 the 5410 and 6560\,\AA\ lines are reproduced very well by
our model. The \ion{C}{iv} 5801/5812\,\AA\ doublet in the synthetic spectrum
is not visible in the observation. It could be that the S/N ratio of the spectrum is not sufficient to resolve the two lines because they are narrower than the \ion{He}{ii} lines. This doublet is known to turn
from absorption into emission at a temperature of about 100\,000\,K \citep{1992LNP...401..273W}. Therefore, a model at the upper limit of our
\Teff\ estimate is in
accordance with the non-detection in the observation. But we note that
at 100\,000\,K the \ion{He}{ii} 4686\,\AA\ line is significantly
weaker and hence gives a poorer fit than our final model with \Teff =
90\,000\,K.

Figure\,\ref{fig:gteff} shows the location of WD1 in the Kiel (\logg\ --
$\log$\,\Teff) diagram together with all other objects of the PG1159 class\footnote{According to an unpublished list based on \cite{2006PASP..118..183W} and maintained by one of the authors (KW).}
and the helium-dominated DO white dwarfs and O(He) stars. WD1 is
remarkable because it is the PG1159 star with the highest surface
gravity. Comparison with the depicted evolutionary tracks gives a mass
of $M = 0.85^{+0.13}_{-0.14}\,M_\odot$, making WD1 the second of the two most
massive members of its spectral class. 

The initial main-sequence progenitor mass $M_{\rm init}$ of a white dwarf in a stellar cluster can be determined from isochrones that are inferred from stellar evolution models. One creates an isochrone at the progenitor's evolutionary lifetime and metallicity. The isochrone's given $M_{\rm init}$ of a star at the tip of the asymptotic giant branch (AGB) is the white dwarf's $M_{\rm init}$. To be consistent with the results of the work of \cite{2018ApJ...866...21C} for other white dwarf members of M37, we adopt a cluster age of $585\pm50$\,Myr and the following procedure to construct PARSEC isochrones \citep{2012MNRAS.427..127B}. We used the web interface CMD3.7\footnote{\url{http://stev.oapd.inaf.it/cgi-bin/cmd_3.7}} and chose PARSEC version 1.2S and a solar metallicity ($Z=0.0152$). The resulting isochrones end at the tip of the AGB. Since the cooling age of WD1 (some $10^5$\,yr; see below) is completely negligible compared to the cluster age, its initial mass is identical to the mass of current M37 stars at the tip of the AGB. We find $M_{\rm init}= 2.77^{+0.09}_{-0.08}$~\Msol, in good agreement with \cite{2022ApJ...935L..35F}. The error reflects the uncertainty of the cluster age. 

\cite{2018ApJ...866...21C} have determined initial and final masses of 14 white dwarfs in M37. They are displayed in Fig.\,\ref{fig:ifmr} (we do not plot an outlier, the ultra-massive white dwarf termed WD33 in their paper) together with WD1, the central star of the PN, which is also the youngest cluster white dwarf. It is an important point in the IFMR because it is independent of any possible uncertainties in white-dwarf cooling theory. The initial and final masses of WD1 are consistent with the IFMR determined by \cite{2018ApJ...866...21C} from eighty white dwarfs, shown by the blue graph in Fig.\,\ref{fig:ifmr}. All these white dwarfs are hydrogen-rich (spectral type DA) in contrast to WD1. However, there is no clear evidence that the IFMR for hydrogen-deficient white dwarfs is different \citep{2021AJ....162..162B}.

\begin{table*}[ht]
    \centering
\caption{Atmospheric and physical parameters of the six DA white dwarfs determined via spectroscopic and photometric analysis.}    
    \begin{tabular}{llllcccc}
    \hline
\noalign{\smallskip}    
      Name ID   & Short Name & {\em Gaia} ID & $T_{\rm eff}$ [K] & $\log{g}$ [cgs] & Mass [M$_\odot$] & $\tau_{\rm cool}$ [Myr] & $d$ [kpc] \\
\noalign{\smallskip}
\hline
\noalign{\smallskip}   
WD\,2 & 0552$+$3236 & {3451182182857026048} & $76000 \pm 4640$ & $8.08 \pm 0.29$ & $0.76 \pm 0.08$ & $0.51^{+0.11}_{-0.21}$ & $1.5^{+0.7}_{-0.5}$ \\
\noalign{\smallskip}   
WD\,3 & 0552$+$3231 & {3451201076423973120} & $94220 \pm 14700$ & $6.91 \pm 0.16$ & $0.52 \pm 0.07$ &  $0.02^{+0.10}_{-0.01}$ & $4.4^{+1.1}_{-0.9}$ \\
\noalign{\smallskip}   
WD\,4 & 0551$+$3216 & {3451167786125150592} & $19080 \pm 1580$ & $7.76 \pm 0.27$ & $0.51 \pm 0.11$ & $112^{+113}_{-45}$ & $0.87 \pm 0.05$\\
\noalign{\smallskip}   
WD\,5 & 0553$+$3229 & {3451200114340263296} & $29630 \pm 1030$ & $8.08 \pm 0.31$ & $0.68 \pm 0.11$ & $11^{+22}_{-2}$ & $0.90 \pm 0.04$\\
\noalign{\smallskip}  
WD\,6 & 0547$+$3246 & {3448258267902375296} & $48690 \pm 1430$ & $7.72 \pm 0.16$ & $0.57 \pm 0.06$ &  $2.4^{+0.3}_{-0.5}$& $1.32^{+0.29}_{-0.24}$\\
\noalign{\smallskip}   
WD\,7 & 0548$+$3323 & {3454340495645123584} & $25420 \pm 1580$ & $8.04 \pm 0.28$ & $0.65 \pm 0.12 $ & $24^{+47}_{-6}$ & $0.76 \pm 0.03$\\
%%%%%%%%%%%%%%%%%%%%%%%%%
%%% Robertos values 
%WD\,3 & 0552$+$3231 & \texttt{3451201076423973120} & $94\,220 \pm 14\,700$ & $6.91 \pm 0.16$ & $0.52 \pm 0.09$ &  $0.02^{+0.19}_{-0.01}$ & $4.4^{+1.1}_{-0.9}$ \\
%\noalign{\smallskip}   
%WD\,5 & 0553$+$3229 & \texttt{3451200114340263296} & $29\,630 \pm 1030$ & $8.08 \pm 0.31$ & $0.68 \pm 0.16$ & $12^{+15}_{-3}$ & $0.90 \pm 0.04$\\
%\noalign{\smallskip}   
%WD\,7 & 0548$+$3323 & \texttt{3454340495645123584} & $25\,420 \pm 1580$ & $8.04 \pm 0.28$ & $0.66 \pm 0.14 $ & $27^{+32}_{-10}$ & $0.76 \pm 0.03$\\
%%%%%%%%%%%%%%%%%%%%%%%%%
\noalign{\smallskip}   
\hline
    \end{tabular}

    \label{tab:da_params}
\end{table*}

\begin{figure}[ht]
\begin{center}
\includegraphics[width=\linewidth]{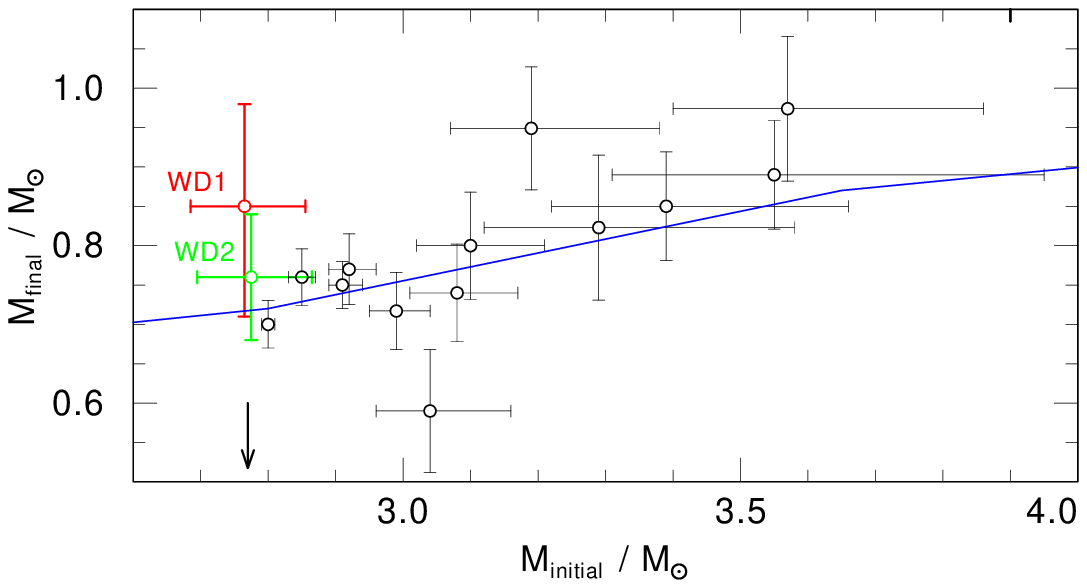}
\caption{
IFMR relation for white dwarfs in the open cluster M37. The red symbol indicates the PN central star WD1 and the green symbol WD2 from our study. The other white dwarfs are from \cite{2018ApJ...866...21C}. The blue graph, taken from that paper,  is the semi-empirical three-piece linear fit to white dwarfs in several open clusters based on PARSEC isochrones. The vertical arrow points to the current M37 progenitor mass at the tip of the AGB.
}
\label{fig:ifmr}
\end{center}
\end{figure}

\begin{figure}[ht]
    \centering
    \includegraphics[width=\linewidth]{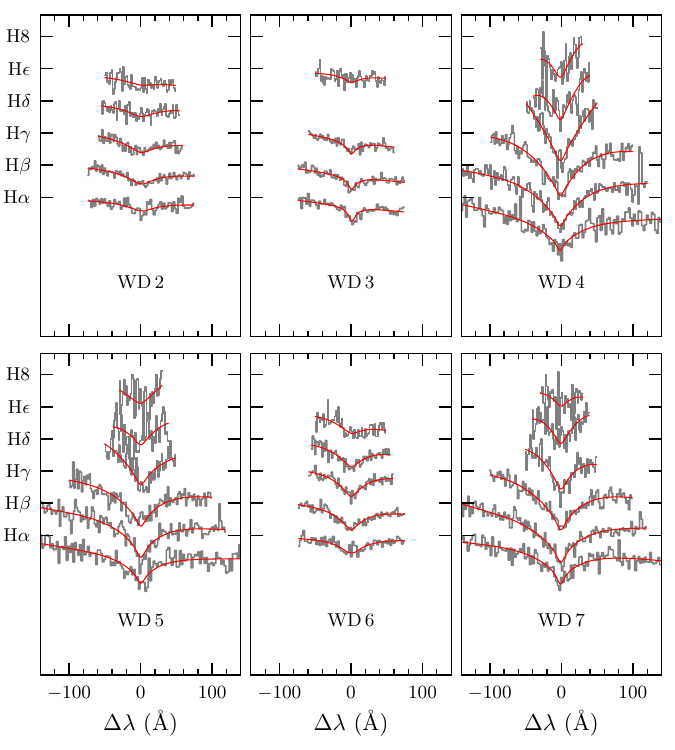}
    \caption{Best-fit results of the spectral analysis for the six DA white dwarfs. The observed Balmer lines and the best-fit models are shown in gray and red, respectively. The respective values for \Teff\ and \logg\ are given in Table~\ref{tab:da_params}.}
    \label{fig:spectral_lines}
\end{figure}

\subsection{DA white dwarfs}
\label{sect:DA}
Using a pure-hydrogen non-LTE model grid computed with \textsc{TMAP} \citep{2023arXiv230703721R}, we derived effective temperatures and surface gravities for the three hottest DA white dwarfs. This was done by fitting the observed Balmer lines by means of $\chi^2$ minimization using the \textsc{FITSB2} routine \citep{Napiwotzki1999} and calculating the statistical 1\,$\sigma$ errors.
For WD3 (WD0552$+$3231) we excluded the H\,$\delta$ line from the fit as it was
affected by a cosmic ray hit. We derive \Teff$=94\,220\pm14\,702$\,K and \loggw{6.91\pm0.16}.
For WD2 (WD0552$+$3236) we find \Teff$=76\,043\pm4640$\,K and \loggw{8.08\pm0.29}, and 
for WD6 (WD0547$+$3246) we determine \Teff$=48\,690\pm1430$\,K and \loggw{7.72\pm0.16}.

The three coolest white dwarfs were analyzed with a grid of hydrogen-dominated synthetic spectra \citep{koester2010}.  We employed FITSB2 for fitting six Balmer lines per star. The best-fit results are shown in Fig.\,\ref{fig:spectral_lines} for both the hot and cool white dwarfs. The atmospheric parameters ($T_{\rm eff}$ and $\log{g}$) are listed in Table\,\ref{tab:da_params}.

\section{SED fitting}
\label{sect:sed}

Subsequently, we  collected available photometry from the Galaxy
Evolution Explorer \citep[GALEX; DR6+7;][]{Bianchi+2017}, IGAPS
\citep{monguio2020}, Pan-STARRS \citep{2016arXiv161205560C}, {\it Gaia} \citep{Gaia+2020},  the XMM-OM Serendipitous Source Survey Catalogue \citep{Page+2012}, and the
Sloan Digital Sky Survey \citep[SDSS; DR12;][]{2015ApJS..219...12A}, in order to estimate the spectrophotometric distances of WD\,1 and the six DA white dwarfs via fits of their SEDs. The $T_{\rm eff}$ and $\log{g}$ are fixed priors, while the white dwarf radii and masses are determined via interpolation from evolutionary tracks. For the DA white dwarfs we employed tracks for He-core and CO-core white dwarfs from \cite{Hall+2013} and \cite{Renedo+2010}, respectively. 
For WD1 we employed very late thermal pulse (VLTP) tracks from \cite{2009ApJ...704.1605A}.
Thus, their distances are estimated through minimization of the $\chi^2$ between observed and synthetic SED, by assuming the radius to be fixed within the measured uncertainties while distance and interstellar extinction are free parameters. The \citet{fitzpatrick2019} extinction law for $R_V = 3.1$ is adopted. For the six DA white dwarfs the results are given in Table\,\ref{tab:da_params}. For WD1 we additionally calculated the radius from the angular diameter (as derived from the SED fit) and the Bailer-Jones distance. Furthermore, we calculated the luminosity from the radius and the effective temperature via $L/L_\odot = (R/R_\odot)^2(T_\mathrm{eff}/T_{\mathrm{eff},\odot})^4$. The results for WD1 are listed in Table\,\ref{tab:resultsall}. In Fig.\,\ref{fig:distances} the spectrophotometric distances of WD1--WD7 are compared to the {\em Gaia}-based distance of M37 obtained by \citet{2022MNRAS.515.1841G}. Their method of SED fitting is the same as ours, so we refrain from showing the SED fits for WD2--WD7. The SED fit for WD1 is shown in Fig.\,\ref{fig:SED}.

One possible explanation for the short spectrophotometric distance determined for WD1 compared with the distance of M37 could be that the optical flux of the model is underestimated because it neglects opacities of all metals except of carbon. They would block the radiation flux in the ultraviolet, which is redistributed into the optical wavelength range. 
To quantify this effect is difficult because the metal abundances can be assessed only from UV spectra. As an example we can look at the DO white dwarf HE\,0504$-$2408. Its atmospheric parameters are similar to WD1: \Teff = 85\,000\,K, \logg = 7. From a UV spectral analysis the metal abundances of eleven species were determined \citep{2018A&A...609A.107W}. Taking the respective model atmosphere and comparing it to a metal-free model shows that the flux increase in the optical by UV metal-line blanketing amounts to 10\%. That would transfer to a distance greater by only 5\% unless the metal abundances are significantly higher.

\begin{figure}[t]
\begin{center}
\includegraphics[width=0.9\linewidth]{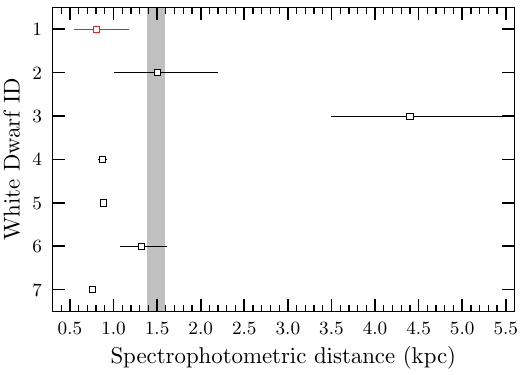}
\caption{Spectrophotometric distances for the six DA white dwarfs and the central star WD1 (red symbol) compared with the {\em Gaia}-based distance of M37 obtained by \citet{2022MNRAS.515.1841G}.}
\label{fig:distances}
\end{center}
\end{figure}

\begin{figure}[t]
\begin{center}
\includegraphics[width=0.9\linewidth]{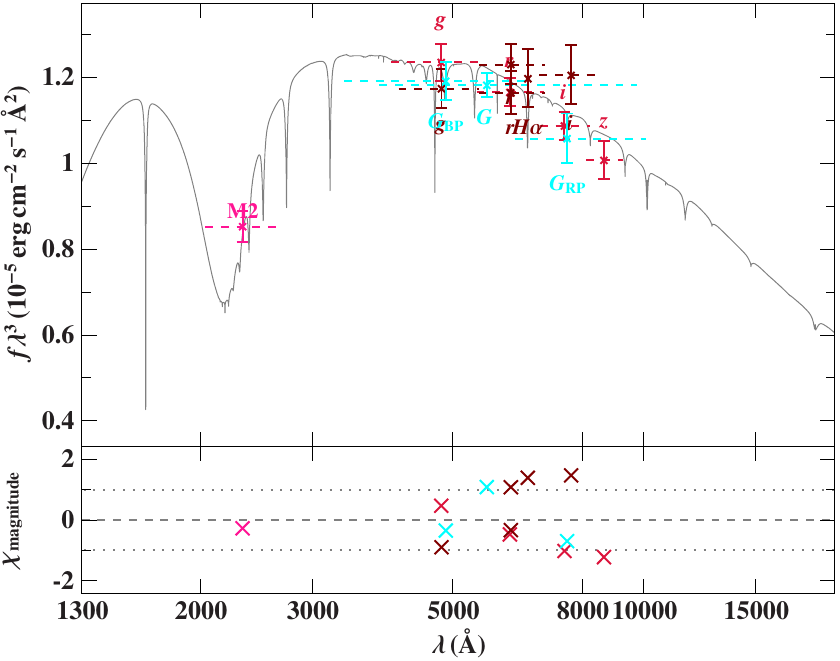}
\caption{SED fit for WD1. Filter-averaged fluxes converted from observed magnitudes are shown in different colors (\textit{Gaia} in cyan, IGAPS in dark red, Pan-STARRS in red, and XMM-OM in magenta). The respective full widths at tenth maximum are shown as dashed horizontal lines. A pure He model with \Teff = 90\,000\,K and \loggw{8.3} that was interpolated from the grid computed in \cite{2023arXiv230703721R} and degraded to a spectral resolution of 6\,\AA\ is plotted in gray. To reduce the steep SED slope, the flux has been multiplied by the wavelength cubed. Bottom panel: Difference between synthetic and observed magnitudes.}
\label{fig:SED}
\end{center}
\end{figure}

\section{Summary and discussion}
\label{sect:discussion}

We have performed a spectroscopic analysis of the central star of the PN IPHASX J055226.2+323724 in the Galactic open cluster M37. The object, WD1, is a hot hydrogen-deficient white dwarf composed of He = $0.94 \pm 0.03$ and C = $0.06 \pm 0.03$ (mass fractions). Its effective temperature of $90\,000 \pm 10\,000$~K and high surface gravity of \logg $= 8.3 \pm 0.3$ in comparison with evolutionary tracks for H-deficient stars \citep{2009ApJ...704.1605A} yield a high mass of $M= 0.85^{+0.13}_{-0.14}$~\Msol. This is significantly higher than the mean mass of white dwarfs \citep[0.61~\Msol;][]{2016MNRAS.455.3413K}. It is also higher than the mass estimated by \cite{2022ApJ...935L..35F}, 0.63~\Msol, from the PN kinematical age and the absolute V magnitude of the PN central star  using evolutionary tracks from \cite{2016A&A...588A..25M}. However, these tracks are for hydrogen-rich stars and therefore not appropriate for WD1.

The post-AGB (cooling) age of WD1 can be estimated from the evolutionary tracks depicted in Fig.\,\ref{fig:gteff}. At the location of WD1, the cooling age is linearly interpolated from the two neighboring tracks, and we determine $\tau_{\rm cool} = 350\,000$~yr. Considering the error bars in temperature and gravity, we find a range of $170\,000-480\,000$~yr. This is significantly longer than the kinematic age of the PN, $78\,000\pm25\,000$~yr, as estimated by \cite{2022ApJ...935L..35F}. The reason for this discrepancy is unclear. They determined the kinematic age assuming a constant PN expansion rate, which might not be realistic. Alternatively, the evolutionary rate of the central star could be underestimated by the stellar models. It was shown recently by \cite{2016A&A...588A..25M} that new calculations with improved input physics can considerably speed up the post-AGB evolution compared to earlier models. However, new models for H-deficient post-AGB stars with masses relevant to WD1 are not available. In any case, because of the high initial mass of the main-sequence progenitor, the PN mass should be rather high, $M\approx 1.9$~\Msol, which would favor a long PN visibility. Only a small fraction of this gas mass is ionized, namely $M=0.32$~\Msol\ \citep{2022ApJ...935L..35F}. Consequently, the real kinematic age of the PN could be even older. In this context we note that there is a faint nebulosity at the southeastern rim of the cluster in Fig.\,\ref{fig:PN} that is not visible in the IPHAS image presented in \cite{2022ApJ...935L..35F}. Spectroscopy and kinematical data could reveal whether it is connected to the PN, although this appears rather unlikely because the PN would have to be exceptionally large and old.

Another fact that complicates the determination of the cooling age of WD1 is its hydrogen-deficient nature. PG1159 stars are assumed to be the result of a late He-shell flash experienced by a post-AGB star or white dwarf that brings the star back onto the AGB \citep[the born-again star scenario; see, e.g.,][]{2006PASP..118..183W}.\footnote{Recently, an alternative channel for the formation of PG1159 stars via a binary white dwarf merger was discussed with the discovery of so-called CO-sdO stars \citep{2022MNRAS.511L..66W}, but this should not produce PG1159 stars associated with a PN.}
Hence, the age derived from the tracks of \cite{2009ApJ...704.1605A} starts from the time when the star left the AGB for the second time. So if we assume that the PN was ejected at the first departure of the star from the AGB, then in principle the PN could be even older. This seems to be highly improbable. Therefore, one can conclude that the late He-shell flash in the precursor of WD1 occurred soon after or even just before the first departure from the AGB; otherwise, the PN would have been dispersed long ago. These are the particular cases of a late thermal pulse (LTP) or AGB final thermal pulse (AFTP) scenarios, respectively, in contrast to the VLTP case that a star suffers as a white dwarf. Evolutionary models predict relatively high amounts of surface hydrogen and nitrogen in the AFTP and LTP cases. So we expect spectral features of these elements in optical spectra of WD1, but their detection would require spectra with higher S/N and resolution.

A further indication that WD1 is a very evolved central star is its carbon abundance. We find C/He = 0.06 (mass ratio). This is the lowest such value of any PG1159 star. C/He = 0.09--1.5 was determined in other members of this spectral class \citep{2006PASP..118..183W}. Due to gravitational settling, carbon will sink out of the atmosphere as soon as the residual radiation-driven wind of the star becomes too weak to be able to act against gravitational settling. This defines the so-called wind limit(s) indicated in the Kiel diagram in Fig.\,\ref{fig:gteff}. A PG1159 star will experience a reduction in carbon when it approaches the wind limit and will eventually transform into a pure-helium DO white dwarf. Wind limits were found from theoretical diffusion models, and their exact location in the Kiel diagram depends on the mass-loss rate assumed for the evolutionary models (which in turn depends, for example, on the metallicity. Figure\,\ref{fig:gteff} shows two such wind limits taken from \cite{2000A&A...359.1042U}. WD1 is very close to these limits and, formally, is the only PG1159 star that has crossed these limits. Obviously, WD1 is in the final stage of PG1159 star evolution and is currently transforming into a DO white dwarf.

The spectrum of WD1 reveals the presence of UHE lines. About 10\% of all hot white dwarfs (\Teff $> 50\,000$~K) show this phenomenon. These lines were tentatively identified as Rydberg
lines of UHE metals (e.g., \ion{O}{VIII}) in ionization stages {\sc V--X} \citep{1995A&A...293L..75W}. This would require a dense environment with temperatures near $10^6$~K, and it has been proposed that the UHE lines form in a
shock-heated, circumstellar magnetosphere \citep{2019MNRAS.482L..93R}. Furthermore, UHE
white dwarfs constitute a new class of variable stars \citep{2021A&A...647A.184R}. Three-quarters of the 16 investigated UHE white dwarfs were found to exhibit a periodic variability with $P=0.22-2.93$~d and amplitudes from a few tenths to a few hundredths of a magnitude. Spots on the surface of these stars and/or geometrical effects of circumstellar material might be responsible for this. Our WD1 central star was classified as a rotating variable with a period of 0.4451~d and an amplitude of 0.074~mag (\citealt{2015AJ....150...27C}; their designation for the star is V1975). Hence, WD1 belongs to this new class of variable UHE stars.

It is also noteworthy that the UHE phenomenon is, for an unknown reason, mainly exhibited by helium-rich white dwarfs (spectral types DO and DOZ; ``Z'' indicates the presence of metal lines, usually from carbon). Only three of the 17 known UHE stars are hydrogen-rich white dwarfs (spectral types DA and DAO; ``O'' indicates the presence of helium). Among the DOZ white dwarfs are two PG1159 stars (i.e., H-deficient objects with very high carbon abundance). WD1 is only the third PG1159 star known to exhibit UHE lines. The other two have temperatures similar to that of WD1. They are SDSS J121523.09$+$120300.8 with \Teff $= 100\,000$~K and \logg $= 7.6$ \citep{2006A&A...454..617H} and WD J070204.29$+$051420.56 with \Teff $= 100\,000$~K and \logg $= 7.5$ \citep{2023arXiv230703721R}.

We determined the main-sequence progenitor mass of WD1, $M_{\rm init}= 2.77^{+0.09}_{-0.08}$~\Msol, from isochrones. WD1 is the youngest member in M37, younger than the 14 white dwarfs studied by \cite{2018ApJ...866...21C}. Its cooling time is negligible compared to the cluster age. Therefore, it is an important point in the IFMR because it is independent of any possible uncertainties in white-dwarf cooling theory. The initial and final masses of WD1 are consistent with the IFMR determined by \citet[see our Fig.\,\ref{fig:ifmr}]{2018ApJ...866...21C}.

\begin{table}[t]
    \centering
\caption{Spectroscopic distance-, proper motion (PM)-, and single star evolution mass-based membership to M\,37 and progenitor parameters of our considered white dwarfs. Proper-motion memberships have been taken from \cite{2023MNRAS.524..108G}. }
    \begin{tabular}{lccc}
     \hline
\noalign{\smallskip}     
Name ID & $d_{\mathrm{spec}}$ & PM & $M\gtrapprox0.7$\Msol \\
\noalign{\smallskip}
     \hline
WD\,1 & \xmark & \cmark & \cmark \\
WD\,2 & \cmark & \cmark & \cmark \\
WD\,3 & \xmark & \cmark & \xmark \\
WD\,4 & \xmark & ? & \xmark \\
WD\,5 & \xmark & \xmark & \cmark \\
WD\,6 & \cmark & ? & \xmark \\
WD\,7 & \xmark & ? & \cmark \\
\hline
    \end{tabular}
    \label{tab:da_progenitors}
\end{table}

Our study reveals that only two of the six DA white dwarfs (WD2 and WD6) have spectrophotometric  distances 
consistent with that of M37 (see Fig.\,\ref{fig:distances} and Table~\ref{tab:da_progenitors}). WD3 appears to be a background star, and the spectroscopic distances of the remaining three DAs suggest that they are foreground stars. Interestingly,  
\cite{2023MNRAS.524..108G} recently found that WD1, WD2, and WD3 should be cluster members according to their proper motions. WD3 is the hottest DA in our sample, and we note that the spectrum of this star was affected by a cosmic particle hit; thus, we had to exclude H$\delta$ from our fit. In addition, it has to be stressed that atmospheric parameters of very hot DA white dwarfs derived from optical
spectra can suffer large systematic errors (e.g., \citealt{Werner+2019}). Therefore, it might be possible that our spectroscopic distance is overestimated. \cite{2023MNRAS.524..108G} find that the proper motions of WD5 are not compatible with those of the cluster. This is consistent with our finding that WD5 is a foreground star. The other white dwarfs (WD4, WD6, and WD7) were not included in the \cite{2023MNRAS.524..108G} study.

%We note that all the studied white dwarfs, but WD1 and WD3, have spectro-photometric distances disagreeing from the Bayesian estimates given by \citet{2021AJ....161..147B}. These estimates assume a smooth exponential prior that accounts for the density of stars along a given Galactic sightline that may not be significant for establishing the membership of these white dwarfs to M37.

Given the current age of M37, the lightest white dwarf members that could have formed through canonical single-star evolution are expected to have masses of $\approx 0.7$\Msol\ \citep{2022MNRAS.515.1841G}. Within the error limits, this is the case for WD1, WD2, WD5, and WD7. However, the mass of WD3 -- whose cluster membership is confirmed according to its proper motion -- is clearly below 0.6\,\Msol. This could be a product of binary evolution, yet it is also possible that our mass is underestimated due to systematic errors in the spectral analysis. The remaining two DAs, WD4 and WD6, have masses of $0.51\pm0.11$\,\Msol\ and $0.57\pm0.06$\,\Msol, respectively. For WD4, a cluster membership seems unlikely in any case due to its spectroscopic distance. But since the spectroscopic distance of WD6 agrees with that of M37, it seems possible that this star evolved through binary evolution. We note that there is an indication that the binary fraction along the main sequence in M37 is close to 20\% \citep{Kalirai+2004}; thus, it appears plausible that at least some of the white dwarfs in M37 have masses below 0.7\,\Msol. In essence, the only DA white dwarf from our sample that is a member of M37 and has a mass consistent with single-star evolution is WD2. Therefore, WD2 can be used to constrain the IFMR, and as such we include it in Fig.\,\ref{fig:ifmr}. As for the central star WD1, its cooling age (0.51\,Myr) is negligible in comparison with the cluster age, and thus both WD1 and WD2 had the same initial mass.

\begin{acknowledgements} 
We thank the referee for a constructive report. NR is supported by the Deutsche Forschungsgemeinschaft (DFG) through grant GE2506/17-1.
MG and LRB acknowledge partial support by MIUR under PRIN programme \#2017Z2HSMF and by PRIN-INAF\,2019.
MEC acknowledges grant RYC2021-032721-I, funded by MCIN/AEI/10.13039/501100011033 and by the European Union NextGenerationEU/PRTR.
PG thanks Marcel Drechsler and Sven Eklund for their contribution to PN imaging. 
RR acknowledges support from Grant RYC2021-030837-I funded by MCIN/AEI/ 10.13039/501100011033 and by “European Union NextGenerationEU/PRTR”. 
This work was partially supported by the AGAUR/Generalitat de Catalunya grant SGR-386/2021 and by the Spanish MINECO grant PID2020-117252GB-I00.
The TMAD tool (\url{http://astro.uni-tuebingen.de/~TMAD})
used for this paper was constructed as part of the activities of the
German Astrophysical Virtual Observatory. 
Based on observations made with the Gran Telescopio Canarias (GTC), installed at the Spanish Observatorio del Roque de los Muchachos of the Instituto de Astrofísica de Canarias, on the island of La Palma.
The Starlink software \citep{berry2022} is currently supported by the East Asian Observatory.
Some of the data presented
in this paper were obtained from the Mikulski Archive for Space
Telescopes (MAST). This research has made use of NASA's Astrophysics
Data System and the SIMBAD database, operated at CDS, Strasbourg,
France. This research has made use of the VizieR catalogue access
tool, CDS, Strasbourg, France. This work has made use of data from the
European Space Agency (ESA) mission {\it Gaia}.

\end{acknowledgements}

\bibliographystyle{aa}
\bibliography{aa}

\end{document}